\documentclass[12pt,letterpaper]{article}
\pdfoutput=1
\usepackage{jheppub}
\usepackage{amsfonts, amsthm}
\usepackage[english]{babel}
\usepackage[utf8]{inputenc}
\usepackage{slashed}
\usepackage{mathrsfs}
\usepackage{amssymb}
\usepackage{color}
\hypersetup{unicode}
\usepackage{natbib}

\newcommand{\eq}{\begin{equation}}
\newcommand{\feq}{\end{equation}}
\newcommand{\be}{\begin{equation}}
\newcommand{\ee}{\end{equation}}
\newcommand{\eqn}{\begin{eqnarray}}
\newcommand{\feqn}{\end{eqnarray}}

\title{$\text{AdS}_D\times I$ solutions in axio-dilaton gravity}

\author[a]{Giuseppe Dibitetto}
\author[b]{Nicol\`o Petri}

\affiliation[a]{Dipartimento di Fisica, Università di Roma “Tor Vergata” \& Sezione INFN Roma 2, Via della
Ricerca Scientifica 1, 00133, Roma, Italy.}
\affiliation[b]{Sezione INFN Milano, Via Celoria 16, 20133, Milano, Italy.}

\emailAdd{giuseppe.dibitetto@roma2.infn.it}
\emailAdd{nicolo.petri@mi.infn.it}

\abstract{We study non-supersymmetric $\mathrm{AdS}_D\times I$ solutions in the context of $(D+1)$ dimensional gravity coupled to an axio-dilaton with arbitrary runaway potential for the dilaton and arbitrary exponential coupling of the dilaton to the axion kinetic energy. We analyze the equations of motion, reformulate them in terms of a first order autonomous dynamical system, and discuss the set of fixed points, their physical interpretation and their stability conditions. We find a few special classes of analytic solutions for arbitrary $D$, including $\mathrm{AdS}_9\times I$ backgrounds in type IIB supergravity. We conclude by discussing the properties of numerical flows including the massive IIA supergravity case.}

\begin{document}
\maketitle
\flushbottom

\section{Introduction}

The existence of non-supersymmetric gravitational vacua remains one of the central open questions in string theory. A particularly influential proposal is the so-called non-SUSY AdS Conjecture of Ooguri and Vafa, according to which there are no stable non-supersymmetric AdS vacua in quantum gravity \cite{Ooguri:2016pdq}. The physical intuition behind this idea is closely related to the Weak Gravity Conjecture \cite{Arkani-Hamed:2006emk}. In the absence of supersymmetry, the balance between gravitational attraction and gauge repulsion is generically lost, implying that the branes supporting the AdS vacuum cannot remain in static equilibrium. As a consequence, non-perturbative brane nucleation processes are expected to destabilize the vacuum.\footnote{For a review on the swampland program, see \cite{Palti:2019pca}.}

While these arguments raise serious doubts about the existence of stable non-supersymmetric AdS vacua, they also leave several important issues unresolved. In fact, even if non-supersymmetric AdS vacua are absent, quantum gravity may still contain a rich spectrum of non-BPS extended objects.

For BPS branes, the situation is relatively well understood. Their near-horizon limit often produces AdS geometries, providing through holography a direct connection between microscopic brane physics and spacetime geometry. Famous examples include M2, D3 and M5 branes, whose near-horizon geometries give rise to some of the best understood realizations of the AdS/CFT correspondence. By contrast, for genuinely non-BPS objects no analogous general picture is currently available. In particular, it remains unclear whether non-supersymmetric extended objects admit an AdS description in the near-horizon limit. Establishing whether such a connection exists would provide valuable insight into the microscopic origin of non-supersymmetric AdS backgrounds and their possible role within quantum gravity.

Recent developments suggest that the Cobordism Conjecture \cite{McNamara:2019rup} provides a particularly promising framework for exploring these questions. In this context, novel non-supersymmetric defects have been proposed in type IIB string theory, most notably the so-called reflection seven-branes, or R7-branes \cite{Dierigl:2022reg,Debray:2023yrs,Dierigl:2023jdp,Chakrabhavi:2025bfi}. These objects are associated with non-trivial axio-dilaton monodromies and appear as natural candidates for genuinely non-BPS defects in string theory. Furthermore, it has been suggested that their local spacetime description, if it exists, may involve codimension-one AdS$_9$ backgrounds \cite{Cavusoglu:2026xiv} (see Appendix D therein). If correct, this would provide a particularly interesting realization of a non-supersymmetric AdS geometry arising from a non-BPS object in the string theory spectrum.

Motivated by these developments, in this work we investigate a general class of AdS$_D \times I$ backgrounds in $(D+1)$-dimensional gravity coupled to an axio-dilaton. Rather than focusing on a specific string-theoretic construction, we adopt a bottom-up perspective and consider the most general class of models involving an exponential runaway potential for the dilaton and an exponential coupling of the dilaton to the axion kinetic term. This class contains a variety of theories arising in supergravity, including the axio-dilaton sector of type IIB supergravity, the dilaton-Romans mass sector of massive IIA supergravity, the eight-dimensional $\mathrm{SU}(2)$ gauged supergravity of Salam and Sezgin, and four-dimensional $\mathcal N=4$ pure supergravity.

Despite the intrinsic difficulty of constructing fully backreacted non-supersymmetric solutions in string theory, significant progress has been achieved in recent years. Notable examples include the construction of fully backreacted AdS$_8$ and dS$_4$ backgrounds in massive IIA supergravity \cite{Cordova:2018eba,Cordova:2018dbb}. Further examples have emerged from the study of bubble solutions and dS-like solutions \cite{Horowitz:2007pr,Ooguri:2017njy,GarciaEtxebarria:2020xsr,Bomans:2021ara,Giri:2021eob,DeLuca:2021pej,Dibitetto:2022rzy,Menet:2025nbf,ValeixoBento:2025yhz,Ghodsi:2026lmn,Andriot:2026lac}, from the framework of Dynamical Cobordisms \cite{Angius:2022aeq,Angius:2023uqk,Huertas:2023syg}, and from the analysis of non-supersymmetric string theories and their associated gravitational backgrounds  \cite{Dudas:2000ff,Mourad:2016xbk,Sagnotti:2021mxb,Raucci:2022bjw,Raucci:2022jgw,Mourad:2024dur,Raucci:2025bev,Basile:2026lyc}.

We study these models by reformulating the second-order equations of motion as a first-order {\itshape autonomous dynamical system}. Our approach follows standard dynamical systems techniques commonly used in cosmology \cite{Copeland:1997et}. The fixed points of this system determine the possible spatial asymptotic regimes of the solutions. We classify these fixed points, analyze their stability properties, and discuss their physical interpretation. In particular, we identify asymptotic regimes that admit a higher-dimensional resolution as well as scaling behaviors with a brane-like interpretation.

We then construct several explicit families of analytic AdS$_D \times I$ solutions. These include backgrounds with a non-trivial axion profile and vanishing scalar potential, relevant for type IIB supergravity, as well as solutions supported by runaway dilaton potentials. Among these solutions are also AdS$_9 \times I$ backgrounds in type IIB, which are studied in detail in the companion paper \cite{Dibitetto:2026oor}. We also discuss numerical flows in cases where analytic solutions are not available. Taken together, our results show that axio-dilaton gravity admits a rich space of non-supersymmetric warped AdS backgrounds.

The paper is organized as follows. In section \ref{Sec:setup} we introduce the general axio-dilaton gravity setup and derive the corresponding equations of motion. We then reformulate the dynamics as a first-order autonomous system and perform a complete analysis of its fixed points and stability properties. In section \ref{Sec:Solutions} we construct explicit classes of AdS$_D \times I$ solutions and discuss their physical interpretation. In appendix \ref{appendix_KK} we discuss a possible higher-dimensional origin of our axio-dilaton models. Finally, in appendix \ref{appendix} we collect the definitions and analytic properties of the hypergeometric functions used in this work.

\section{Gravity in $(D+1)$ dimensions coupled to an axio-dilaton}
\label{Sec:setup}

Let us first introduce our setup. We consider Einstein gravity in $(D+1)$ dimensions coupled to an axio-dilaton, with $D>1$. Our interest in this class of theories is motivated by the following reason. This is the same setup considered in \cite{Cavusoglu:2026xiv}, where it is argued that codimension 2 objects with exotic monodromies should exist in the quantum spectrum of the type IIB string, \emph{a.k.a.} R7 branes. These objects are solutions of axio-dilaton gravity and similar solutions are argued to exist in arbitrary spacetime dimensions. In \cite{Cavusoglu:2026xiv} it is also observed that the metric solutions associated to R7 branes admit a limit where the geometry becomes $\mathrm{AdS}_9$ times an interval. By invoking the holographic correspondence, one might then argue for the existence of strongly coupled conformal phases in non-supersymmetric higher dimensional QFT's. Therefore, this seems to be the perfect playground to investigate non-supersymmetric AdS solutions and their dynamical brane origin, which is to our knowledge, still an urgent open problem. The classical action reads
\begin{equation}
S_{D+1} \ = \ \frac{1}{2\kappa_{D+1}^2}\,\int d^{D+1}x \sqrt{-g_{D+1}}\,\left(\mathcal{R}_{D+1}\,-\,\frac{\partial \tau \partial \bar{\tau}}{2\mathrm{Im}(\tau)^2}\,-\, V(\tau,\bar{\tau})\right) \ ,
\end{equation}
where the complex axio-dilaton is defined as $\tau\,\equiv\, \lambda \,+\, i\,e^{-\phi}$, with $\lambda$ \& $\phi$ being real fields. In terms of these real dof's, the above action may be recast in a slightly more general form
\begin{equation}\label{action}
S_{D+1} \ = \ \frac{1}{2\kappa_{D+1}^2}\int d^{D+1}x \sqrt{-g_{D+1}}\,\left(\mathcal{R}_{D+1}-\frac{1}{2}(\partial\phi)^2-\frac{1}{2}f(\phi)^2(\partial\lambda)^2- V(\phi)\right) \ ,
\end{equation}
where we shall assume the axion kinetic coupling and the scalar potential to take the following form
\begin{equation}\label{alpha_beta}
\begin{array}{lcccclc}
f(\phi) \ = \ e^{\beta\phi} & , & & \textrm{and} & & V(\phi) \ = \ m^2\, e^{2\alpha\phi} & ,
\end{array}
\end{equation}
where $\alpha$, $\beta$ and $m^2$ are arbitrary real parameters, while we still keep the possibility open to cases where $m^2<0$. We refer to appendix \ref{appendix_KK} for a complete discussion of the possible higher dimensional origin of Lagrangians of the form \eqref{action}, through dimensional reductions over an arbitrary Einstein manifold of gravity coupled to a $d$-form gauge potential completely wrapping internal space. When it comes to possible stringy features of these models, a few special cases of interest that are included in our setup are summarized in table \ref{table:special_cases}.
\begin{table}[htp]
\begin{center}
\begin{tabular}{|c|c|c|c|}
\hline
$\left(D,\alpha,\beta\right)$ & $m^2$ & Axion? & Theory   \\[2mm]
\hline \hline
$\left(9,\frac{5}{4},-\right)$ & $>0$ & No & massive IIA supergravity  \cite{Romans:1985tz}\\
\hline
$\left(9,-,1\right)$ & $=0$ & Yes & IIB supergravity  \\
\hline
$\left(7,\frac{1}{2},1\right)$ & $<0$ & Yes & 8D $\mathrm{SU}(2)$ gauged supergravity  \cite{Salam:1984ft}\\
\hline
$\left(3,-,1\right)$ & $=0$ & Yes & 4D $\mathcal{N}=4$ pure supergravity  \\
\hline
\end{tabular}
\end{center}
\caption{\it A few interesting particular cases included in our setup. As mentioned earlier, the runaway potential in the dilaton is sometimes positive, as \emph{e.g.} in massive IIA supergravity, or negative, as in gauged supergravities with compact gauge groups.}
\label{table:special_cases}
\end{table}

By varying the action \eqref{action} w.r.t. the $(D+1)$ dimensional metric and the axio-dilaton, we find the following field equations
\begin{eqnarray}
G^{(D+1)}_{\mu\nu}+\frac{1}{2}V g_{\mu\nu} & = & \frac{1}{2}\left(\partial_{\mu}\phi\partial_{\nu}\phi- \frac{1}{2}(\partial\phi)^2g_{\mu\nu}\right)+\frac{f^2}{2}\left(\partial_{\mu}\lambda\partial_{\nu}\lambda- \frac{1}{2}(\partial\lambda)^2g_{\mu\nu}\right) \ , \nonumber \\
\partial_{\mu}\left(\sqrt{-g_{D+1}}\,\partial^{\mu}\phi\right) & = & \sqrt{-g_{D+1}}\left(f\,f_{\phi}(\partial\lambda)^2+V_{\phi}\right) \ , \label{General_EOMs}\\
\partial_{\mu}\left(\sqrt{-g_{D+1}}\,f^2\partial^{\mu}\lambda\right) & = & 0 \ , \nonumber 
\end{eqnarray}
where $G^{(D+1)}_{\mu\nu}$ denotes the $(D+1)$ dimensional Einstein tensor, while $f_{\phi}$ \& $V_{\phi}$ denote the $\phi$ derivatives of $f$ and $V$, respectively.
 
The general form of the metric we choose is that of a codimension one AdS slicing
\begin{equation}
ds^{2}_{D+1} \, = \, e^{2A(y)}L^{2}ds_{\mathrm{AdS}_{D}}^{2} \, + \, e^{2B(y)} dy^2\ ,
\end{equation}
where the metric for the $D$ dimensional slices $ds_{\mathrm{AdS}_{D}}^{2}$ is that of unit AdS, and the warp factors $A$ \& $B$, along with $\lambda$ \& $\phi$ are assumed to have a profile in $y$. It is worth noticing that $B$ is strictly speaking \emph{non-dynamical}, in the sense that its second derivatives never appear in the field equations. This is due to the fact that $B$ may always be reabsorbed into a reparametrization of the $y$ coordinate. Nevertheless, keeping this gauge freedom may prove to be very useful when searching for analytic solutions to the field equations. By adapting the field equations \eqref{General_EOMs} to our \emph{Ansatz}, we get the following system of coupled second order ODE's
\begin{eqnarray}
\hspace{-2mm}0&=&\frac{(D-1)(D-2)}{2L^2} e^{2(B-A)}+(D-1)\left(A''-A'B'+\frac{D}{2}(A')^2\right)+\frac{(\phi')^2+f^2(\lambda')^2}{4}+\frac{V}{2}e^{2B}\ , \nonumber \\
\hspace{-2mm}0&=&\left(e^{DA-B}\phi'\right)' -e^{DA-B}\left(f\,f_{\phi}(\lambda')^2+e^{2B}V_{\phi}\right) \ , \label{D_EOMs}\\
\hspace{-2mm}0&=&\left(e^{DA-B}f^2\lambda'\right)'  \ , \nonumber 
\end{eqnarray}
where $'$ denotes differentiation w.r.t. $y$. As usual, the above second order system is supplemented with a \emph{Hamiltonian constraint} coming from the Einstein equation along $yy$, which reads
\begin{equation}\label{HCD}
\frac{D(D-1)}{2L^2} e^{2(B-A)}+\frac{D(D-1)}{2}(A')^2-\frac{(\phi')^2+f^2(\lambda')^2}{4}+\frac{V}{2}e^{2B}=0 \ .
\end{equation}
This constraint has identically vanishing $y$ derivative on-shell, by virtue of the second order field equations \eqref{D_EOMs}. This second order dynamics may be derived from the following 1D effective Lagrangian
\begin{equation}
L_{\textrm{eff}} \ = \ e^{DA-B}\left( D(D-1) (A')^2\,-\,\frac{1}{2}\left((\phi')^2+f^2(\lambda')^2\right)\right)-e^{DA+B}\left(V+\frac{D(D-1)}{L^2}e^{2A}\right) \ .
\end{equation}
In that case, the associated Hamiltonian vanishes identically on-shell as a consequence of the constraint \eqref{HCD}, whence its name.

\subsection{The 1D autonomous system description and fixed point analysis}
The above system of second order field equations may be reformulated in terms of a 1D autonomous system, following \cite{Copeland:1997et} as it is commonly done in the contetxt of time flows in cosmology (see \emph{e.g.} \cite{Brinkmann:2022oxy} for an application to multifield quintessence). 
Applying the same logic \emph{mutatis mutandis}, we introduce the following state variables
\begin{equation}\label{XYZ_def}
\begin{array}{lccclccclc}
X \ \equiv \ \gamma_D^{-1}\,\dfrac{\phi'}{A'}&  & , & & Y \ \equiv \ \gamma_D^{-1}\,\dfrac{f\lambda'}{A'} & & , & & Z \ \equiv \ \gamma_D^{-1}e^{B}\,\dfrac{\sqrt{2V}}{A'} & ,
\end{array}
\end{equation}
where we also introduced the shorthand notation $\gamma_D \,\equiv\, \sqrt{2D(D-1)}$. Two comments are due at this point. Firstly, we should remark that the above objects are gauge invariant, in the sense that they are independent of the choice of parametrization for the $y$ line. Secondly, for the way we defined them, $X$, \& $Y$ are manifestly real, while $Z$ might be \emph{imaginary} if $V<0$.

The Hamiltonian constraint \eqref{HCD} may be then rewritten in terms of $(X,Y,Z)$ as
\begin{equation}\label{HCD_XYZ}
X^2\,+\,Y^2\,-\,Z^2 \ = \ 1 \,+\, \frac{e^{2(B-A)}}{L^2(A')^2} \ ,
\end{equation}
which is to be interpreted as a sort of energy balance identity that must be respected by any dynamical flow. In particular, the following special cases may be treated in this framework as constrained systems where the $(X,Y,Z)$ coordinates are no longer left free to vary in $\mathbb{R}^3$, but rather they are confined on a surface.
\begin{itemize}
\item Type IIB-like scenarios (\emph{i.e.} $m^2=0$): restrict to the $Z=0$ plane,
\item Massive type IIA-like scenarios (\emph{i.e.} $\lambda\equiv 0$): restrict to the $Y=0$ plane,
\item Flat slicing (\emph{i.e.} $L=\infty$): restrict to the $X^2+Y^2-Z^2=1$ hyperboloid.
\end{itemize}
The full set of second order field equations \eqref{D_EOMs} is automatically implied by the following set of first order flow equations
\begin{eqnarray}
\frac{dX}{dA} & = & DX\left(X^2+Y^2-1\right)+\gamma_D\left(\frac{f_{\phi}}{f}Y^2+\frac{1}{2}\frac{V_{\phi}}{V}Z^2\right)-X\left(X^2+Y^2-Z^2-1\right)\ ,\nonumber\\
\frac{dY}{dA} & = &  DY\left(X^2+Y^2-1\right)-\gamma_D\frac{f_{\phi}}{f}XY-Y\left(X^2+Y^2-Z^2-1\right)\ , \label{XYZ_flow}\\
\frac{dZ}{dA} & = & DZ\left(X^2+Y^2\right)+2\gamma_D\frac{V_{\phi}}{V}XZ-Z\left(X^2+Y^2-Z^2-1\right)\ .\nonumber
\end{eqnarray}
It may be worth mentioning that the above flow is naturally expressed in a gauge fixed form, \emph{i.e.} where we use $A$ as a radial coordinate. However, this may be translated into an arbitrary gauge by simply scaling all the lhs's of \eqref{XYZ_flow} by an appropriate function of the new radial coordinate. Unfortunately, it does not appear obvious how to make use of this freedom in general to analytically integrate the flow equations  for arbitray $(D,f(\phi),V(\phi))$. Since this option seems to be out of reach for the moment, we will start by performing a study of the fixed points of the system, that may provide clues on the qualitative properties of the existing different flows in the various cases.

Fixed points (FP) for this dynamical system are defined as configurations with constant $(X,Y,Z)$. Hence, in order to fully classify the FP in our case, we just need to solve an algebraic system of equations. These become
\begin{eqnarray}
0 & = & DX\left(X^2+Y^2-1\right)+\gamma_D\left(\beta Y^2+\alpha Z^2\right)-X\left(X^2+Y^2-Z^2-1\right)\ ,\nonumber\\
0 & = & Y\left( (D-1)\left(X^2+Y^2-1\right)+Z^2-\gamma_D\beta X\right)\ ,\\
0 & = & Z\left( D\left(X^2+Y^2\right)-\left(X^2+Y^2-Z^2-1\right)+\gamma_D\alpha X\right)\ ,\nonumber
\end{eqnarray}
where we have used \eqref{alpha_beta}. An exhaustive classification of the different FP of the system may be found in table \ref{table:FP}.
\begin{table}[htp]
\begin{center}
\begin{tabular}{|c|c|c|}
\hline
Family Id & $\left(X_0,Y_0,Z_0\right)$ & Extra labels  \\[2mm]
\hline \hline
I & $\left(\sigma,0,0\right)$ & $\sigma\in\mathbb{Z}_2$\\
\hline
II & $\left(-\frac{D}{\gamma_D(\alpha+\beta)},\sigma_1\sqrt{\frac{\alpha}{\alpha+\beta}-\frac{D^2}{\gamma^2_D(\alpha+\beta)^2}},\sigma_2\sqrt{\frac{-\beta}{\alpha+\beta}}\right)$ &  $\left(\sigma_1,\sigma_2\right)\in\mathbb{Z}_2\times\mathbb{Z}_2$\\
\hline
III & $\left(-\frac{1}{\gamma_D\alpha},0,\sigma\frac{i}{\sqrt{2D}\alpha}\right)$ & $\sigma\in\mathbb{Z}_2$\\
\hline
IV & $\left(-\frac{\gamma_D\alpha}{D},0,\sigma\sqrt{\frac{\gamma^2_D\alpha^2}{D^2}-1}\right)$ & $\sigma\in\mathbb{Z}_2$\\
\hline
\end{tabular}
\end{center}
\caption{\it An exhaustive classification of the different FP of the system. Each family depends on additional discrete labels $\sigma_i$ that may be $\pm 1$. Note that family II only exists when $\alpha(\alpha+\beta)\geq D^2\gamma_D^{-2}$, which guarantees a real value for $Y$.}
\label{table:FP}
\end{table}

\subsection{Physical interpretation of the different families of FP}
After identifying the different FP of the system that may or may not occur depending on $(D,\alpha,\beta)$, we would like to physically interpret their meaning. We remind readers that having constant values for $(X,Y,Z)$ does not \emph{per se} imply a constant radial profile for the physical fields. Rather, these field profiles will represent special asymptotic configurations that may be obtained by integrating the field equations in particular limits where some contributions to the effective Hamiltonian are assumed to be dominant, while others are simply disregarded.
\begin{itemize}
\item \textbf{Family I:} $X=\pm 1$ implies $\phi'\,=\,\pm \gamma_D A'$ and hence, up to shifts, $\phi\,=\,\pm \gamma_D A$. By checking the Hamiltonian constraint in the form of \eqref{HCD_XYZ}, we realize that the corresponding field configuration is an asymptotic solution valid in the flat slicing limit.
With this precise ratio, the $(D+1)$ dimensional metric and the dilaton exactly combine into the flat $(D+2)$ dimensional metric written as $\mathrm{Mkw}_{D}\times\mathbb{R}^2$. This implies that even though the $(D+1)$ dimensional metric and the dilaton are \emph{singular}, this singularity gets smoothed out in one higher dimension. Thus, we expect any flow ending up in a FP of this type to have a finite on-shell Euclidean action.

\item \textbf{Family II:} It is an asymptotic scaling solution, which is valid in the flat slicing limit. This field configuration reads
\begin{equation}
\begin{array}{lclclc}
e^{B} \,=\, e^{\frac{\alpha D}{\alpha+\beta}A} & , & e^{\phi} \,=\, e^{-\frac{D}{\alpha+\beta}A} & ,  & \lambda \,=\, \lambda_0 e^{\frac{\beta D}{\alpha+\beta}A} & ,
\end{array}
\end{equation}
where $\lambda_0$ is fixed by the Hamiltonian constraint.
A clear interpretation of this is available, at least for those $(\alpha,\beta)$ complying with the constraints in \eqref{KK_dictionary}, hence admitting a higher dimensional description. By using the uplift formula in \eqref{KK_Ansatz}, one finds a $(D+d+1)$ dimensional background of the form
\begin{equation}
\begin{array}{lclc}
ds_{D+d+1}^2 & = & e^{2\left(1-\frac{\xi D}{\alpha+\beta}\right)A} ds_{\mathrm{Mkw}_D}^2\,+\, e^{\frac{2D(\alpha-\xi)}{\alpha+\beta}A} dA^2 \, + \, e^{-2\frac{\zeta D}{\alpha+\beta}A}ds_{\mathcal{M}_d}^2 & , \\[2mm]
C_{(d)} & = & \lambda_0 \,e^{\frac{\beta D}{\alpha+\beta}A} \mathrm{vol}_{\mathcal{M}_d} & ,
\end{array}
\end{equation}
where $\xi$ \& $\zeta$ are fixed by the condition \eqref{xizeta}, and $\mathcal{M}_d$ is an Einstein manifold that we can think of as being $S^d$, for simplicity. This structure exactly fits the form of the near-horizon limit of an extremal $(D-1)$ brane which is magnetically coupled to the $d$-form gauge potential $C_{(d)}$
\begin{equation}
\begin{array}{lclc}
ds_{D+d+1}^2 & = & H^{-\frac{2}{D}} ds_{\mathrm{Mkw}_D}^2\,+\, H^{\frac{2}{d-1}}  ds_{\mathbb{R}^{d+1}}^2 & , \\[2mm]
C_{(d)} & = & \lambda_0 \,H^{-1} \mathrm{vol}_{S^d} & ,
\end{array}
\end{equation}
where $H\,\sim\,r^{-(d-1)}$, once the appropriate radial coordinate is introduced.
\item \textbf{Family III:} This FP corresponds with a $A=B=- \alpha\phi$ solution, which is real only if both $m$ and $L$ are imaginary, \emph{i.e.} $\mathrm{AdS}_{D}\rightarrow \mathrm{dS}_{D}$ and $m^2<0$. The Hamiltonian constraint then fixes $L$ to be
\begin{equation}\label{m2L2}
m^2L^2 \,=\, \frac{D-1}{2(D-1)\alpha^2-1} \ .
\end{equation}
At least for those models that admit a higher dimensional origin in the sense explained in appendix \ref{appendix_KK}, we can interpret this family of FP as nothing but higher dimensional flat space. Indeed, one can rewrite $\mathrm{Mkw}_{D+d+1}$ as \cite{Dibitetto:2020csn}.
\begin{equation}
ds_{\mathrm{Mkw}_{D+d+1}}^2 \,=\, dr^2+L^2r^2 ds_{\mathrm{dS}_{D}}^2 +R^2r^2ds_{S^{d}}^2 \ ,
\end{equation}
with the constraint $\frac{L}{R} = \sqrt{\frac{D-1}{d-1}}$.
By using \eqref{KK_dictionary} inside our \eqref{m2L2} to write $m$ \& $\alpha$ in terms of $(R,D,d)$, we exactly obtain the aforementioned required   $\frac{L}{R}$ ratio, as a consequence of the Hamiltonian constraint.

\item \textbf{Family IV:} It is again a solution in the flat slicing limit. This time it corresponds to an end-of-the-world (ETW) brane of the type discussed in \cite{Angius:2022aeq}. It is a flat DW background sourced by the runaway potential parametrized by $m^2$, which reads
\begin{equation}
\begin{array}{lcl}
ds_{D+1}^2 & = & H^{\tau_D/D}ds^2_{\textrm{Mkw}_D}\,+\, H^{\tau_D}dz^2 \ , \\
e^{\alpha\phi} & = & H^{-\left(1+\tau_D/2\right)} \ ,
\end{array}
\end{equation}
where $\tau_D\equiv 2 \left(\left(\dfrac{\alpha}{\alpha_{*}^{(D)}}\right)^2-1\right)^{-1}$, with $\alpha_{*}^{(D)}\equiv D\gamma_D^{-1}$, and $\dfrac{dH}{dz}\,=\, \sqrt{2}m$. In order to check our statement, we simply need to evaluate 
\begin{equation}
\frac{\phi'}{A'}\,=\, \frac{\frac{d\phi}{dz}}{\frac{dA}{dz}}\,=\, -\frac{D\alpha}{{\alpha_{*}^{(D)}}^2}\,=\, -\gamma_D^2\frac{\alpha}{D} \ ,
\end{equation}
exactly as it should. It is worth mentioning that this FP, when $D=9$ and $\alpha=\frac{5}{4}$, corresponds to ending in a D8/O8 singularity, depending on the sign of $m$.
\end{itemize}

\subsection{Stability analysis and phase diagram}
Studying the stability conditions for a given FP means to analyze the evolution of the fluctuations around it along the flow. Let $\textbf{R}\equiv(X,Y,Z)$ and let $\textbf{R}_0\equiv(X_0,Y_0,Z_0)$ be a FP, \emph{i.e.} such that $\textbf{F}(\textbf{R}_0)=\textbf{0}$, where we denoted\footnote{This notation should remind us that $\textbf{F}$ is essentially a generalized force field, which is in our case non-conservative, since it cannot be expressed as the gradient of a potential.} by \textbf{F}(\textbf{R}) the rhs's of \eqref{XYZ_flow} as an $\mathbb{R}^3$ valued function of $(X,Y,Z)$. The radial evolution of small fluctuations around $\textbf{R}_0$ obey
\begin{equation}
\left(\textbf{R}-\textbf{R}_0\right)^{\boldsymbol{\cdot}} \,=\, \underbrace{\textbf{F}(\textbf{R}_0)}_{\textbf{0}} \,+\, J_{\textbf{F}}(\textbf{R}_0)\left(\textbf{R}-\textbf{R}_0\right) \,+\, \dots \ ,
\end{equation}
where $J_{\textbf{F}}(\textbf{R}_0)$ denotes the Jacobian matrix of $\textbf{F}$, and $\boldsymbol{\cdot}$ denotes a derivative w.r.t. $A$. Fluctuations are then naturally attracted back to the FP if $J_{\textbf{F}}(\textbf{R}_0)$ has only {negative eigenvalues}. However, we would like to remark that this conclusion turns out to be a bit too naive in our situation. This misunderstanding stems from using $A$ as a radial variable. Indeed, one expects dynamical flows to interpolate between pairs of FP $\left(\textbf{R}_1,\textbf{R}_2\right)$, the former being unstable and the latter being stable. Still, this would only straightforwardly connect to the sign of the eigenvalues of $ J_{\textbf{F}}$ only if $A$ grew monotonically along the flow. On the contrary, all solutions studied in this paper have an inversion point where $A'=0$, and $A'$ has different signs on the two sides, much like in a Euclidean wormhole geometry. This in turn implies that good FP that may behave as attractors in this case are rather those ones characterized by a Jacobian matrix with only \emph{positive eigenvalues}. We shall denote these by stable FP in our work. Furthermore, even if a FP is not fully stable in the sense we just introduced, it might still play the role of an attractor in some flows, provided that the solution spontaneously evolve as to dodge the unstable directions in the asymptotic regions. A similar phenomenon is already well known to occur in the context of static flat DW geometries dual to field theory RG flows.

Let us now examine the stability condition for the four different families.
\begin{itemize}
\item \textbf{Family I:} This family depends on a single sign $\sigma=\pm 1$. The $\sigma$ dependent expression for the eigenvalues of the Jacobian reads
\begin{equation}
\mathrm{Eig}\left(J_{\textbf{F}}(\textbf{R}_0)\right)=\left\{2(D-1);\  D+\sigma\gamma_D\alpha;\ -\sigma\gamma_D\beta\right\}\ .
\end{equation}
This implies stability if
\begin{equation}
\begin{array}{lcccccl}
\left\{\begin{array}{l}
\sigma =+1 \ , \\
\alpha > -D\gamma_D^{-1} \ ,\\
\beta < 0 \ ,
\end{array}
\right.
& & & \textrm{or} & & &
\left\{\begin{array}{l}
\sigma =-1 \ , \\
\alpha < D\gamma_D^{-1} \ ,\\
\beta > 0 \ .
\end{array}
\right.
\end{array}
\end{equation}

\item \textbf{Family II:} In this case a general closed form expression for the eigenvalues in terms of $(D,\alpha,\beta,\sigma_1,\sigma_2)$ is not available. Generically, we observe that these are not even guaranteed to be real. A relevant case with complex eigenvalues actually happens to occur for the Salam \& Sezgin theory (SU$(2)$ gauged 8D supergravity, row 3 in table \ref{table:special_cases}), \emph{i.e.} for $D=7, \ \alpha=\frac{1}{2}$ and $\beta=1$. An interesting case with real eigenvalues is instead the one of massive IIA-like scenarios with arbitray $\alpha$, but $\beta=0$.  In this case, the eigenvalues of the Jacobian are universally given by $\left\{2(D-1);\  0;\ 0\right\}$, but only for $\alpha\geq\sqrt{\frac{D}{2(D-1)}}$ the associated FP has a real $Y$. For those $\alpha$'s this is a candidate attractor point.

\item \textbf{Family III:} Independently of $\sigma=\pm 1$, the expression for the eigenvalues of the Jacobian reads
\begin{equation}
\mathrm{Eig}\left(J_{\textbf{F}}(\textbf{R}_0)\right)=\left\{\frac{\beta}{\alpha}+1-D;\  -\frac{D-1}{2}\pm\sqrt{\frac{1}{\alpha^2}+\frac{(D-1)(D-9)}{4}}\right\}\ .
\end{equation}
An immediate consequence is that FP in this family are always saddles.

\item \textbf{Family IV:} The expression for the eigenvalues turns out again to be $\sigma$ independent. These read
\begin{equation}
\mathrm{Eig}\left(J_{\textbf{F}}(\textbf{R}_0)\right)=\left\{-D\left(1-\frac{\alpha^2\gamma_D^2}{D^2}\right); -2\left(1-\frac{\alpha^2\gamma_D^2}{D}\right);  -D\left(1-\frac{\alpha(\alpha+\beta)\gamma_D^2}{D^2}\right)\right\}\ .
\end{equation}
This implies stability if
\begin{equation}
\begin{array}{lcccccl}
\left\{\begin{array}{l}
\alpha < -D\gamma_D^{-1} \ ,\\
\beta < -\alpha\left(1-\frac{D^2}{\gamma_d^2\alpha^2}\right) \ ,
\end{array}
\right.
& & & \textrm{or} & & &
\left\{\begin{array}{l}
\alpha > D\gamma_D^{-1} \ ,\\
\beta > -\alpha\left(1-\frac{D^2}{\gamma_d^2\alpha^2}\right) \ .
\end{array}
\right.
\end{array}
\end{equation}
\end{itemize}

\begin{figure}[!http]
	\centering
	\includegraphics[width=0.75\textwidth]{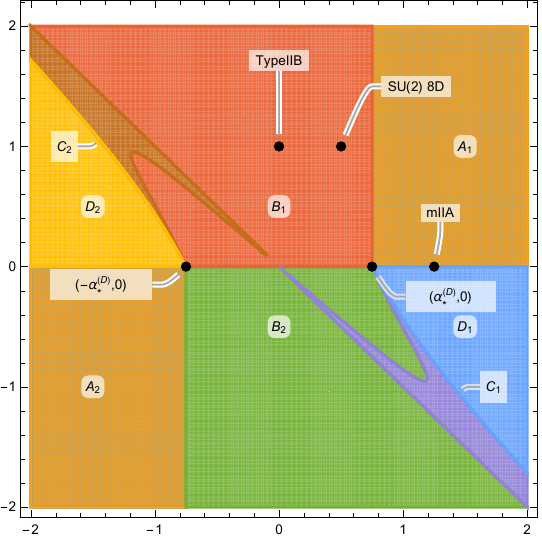}
	\caption{\it  The phase diagram of our 1D system. Each point in this plane represents a specific choice of $(\alpha,\beta)$, while $D$ is kept generic, even though for the sake of concreteness, we drew it for $D=9$. Regions A have FP IV as the only stable ones, regions B have only I, regions C have I and II, regions D have I and IV.
	}
	\label{fig:Phase_Diagram}
\end{figure}

We are now finally ready to draw a qualitative phase diagram, whose shape is universal for any $D$. At fixed $D$, each point in the $(\alpha,\beta)$ plane represents a concrete model. A multiple point in this diagram is represented by $(\alpha_*^{(D)},0)$, where the critical value $\alpha_*^{(D)}$ is given by
\begin{equation}\label{alpha_crit}
\alpha_*^{(D)} \ = \ D\gamma_D^{-1} \ = \ \sqrt{\frac{D}{2(D-1)}} \ .
\end{equation}
A detailed sketch of the phase diagram for our 1D autonomous system may be found in figure \ref{fig:Phase_Diagram}. The diagram is manifestly symmetric w.r.t. to flipping $(\alpha,\beta) \ \longrightarrow \ (-\alpha,-\beta)$, which simply amounts to sending $\phi$ into $-\phi$.

\section{Explicit analytic solutions}\label{Sec:Solutions}
Let us now concentrate on the different classes of analytic solutions to the field equations \eqref{D_EOMs} that we were able to find. In general, a key ingredient that allows one to integrate the equations is the use of the gauge freedom in parametrizing the $y$ coordinate through an appropriate selection of the function $B$ appearing in the $(D+1)$ dimensional metric. Our analytic solutions for arbitrary $D$ are mainly organized in three different groups.
\begin{itemize}
\item $\alpha=\frac{1}{2}$, $\beta=1$ with $m^2<0$, including the maximal $\mathrm{SU}(2)$ gauged 8D supergravity (when $D=7$)

\item No potential ($m^2=0$), arbitrary $\beta$, including type IIB supergravity when $\beta=1$ and $D=9$

\item No axion ($\lambda\equiv 0$), $m^2>0$, for $\alpha \,=\, \dfrac{1}{\gamma_D}; \ \dfrac{D+1}{2\gamma_D}; \ \dfrac{D-2}{\gamma_D}$
\end{itemize}

\subsection{The $\alpha=\frac{1}{2}$, $\beta=1$ family and first order flows}
For these special values of $\alpha$ \& $\beta$, one may observe that the energy balance of the system splits into two independent sectors, which separately take a fraction $p\in (0,1)$ and $(1-p)$ of the total energy, respectively. As a consequence, the following two Hamiltonian constraints
\begin{equation}
\left\{
\begin{array}{rclc}
p\dfrac{D(D-1)}{2} (A')^2+ (1-p)\dfrac{D(D-1)}{2L^2}e^{2 (B- A)}-\dfrac{1}{4} \left((\phi')^2+e^{2 \phi} (\lambda ')^2\right) & = & 0 & ,\\[2mm]
 (1-p) \dfrac{D(D-1)}{2} (A')^2+p\dfrac{D(D-1)}{2L^2}e^{2 (B- A)}-\dfrac{g^2}{2} e^{2 B+\phi}  & = & 0 & ,
\end{array}\right.
\end{equation}
hold on-shell, provided that $p\overset{!}{=}\frac{D-1}{2D}$. In the above, we set $m=i g$, where $g$ is real.
These, together with $\lambda' \sim (\textrm{const}) e^{-DA+B-2\phi}$, define the following first order flow
\begin{equation}
\hspace{-3mm}
\left\{
\begin{array}{lclc}
A' & = & \dfrac{e^{B-A}}{L}{ \sqrt{\dfrac{2g^2 L^2  e^{2 A+\phi }-(D-1)^2}{D^2-1}}} & ,\\[3mm]
\phi' & = & (D-1)  \dfrac{e^{-3 A+B-\phi}}{ g^2L^3} \sqrt{\dfrac{2g^2 L^2  e^{2 A+\phi }-(D-1)^2}{D^2-1}} \left(g^2L^2  e^{2 A+\phi }+2(D-1)\right) & ,\\[3mm]
\lambda' & = &{\sqrt{\dfrac{D-1}{D+1}} \dfrac{e^{-3 A+B-2\phi}}{ g^2L^3} \left( (D-3) g^2 L^2e^{2 A+\phi }+2 (D-1)^2\right)} & ,
\end{array}\right.
\end{equation}
which directly implies the field equations. This flow may be easily integrated by choosing $B=A$ as a gauge, in such a way that the flow equations for $A$ \& $\phi$ combine into a single equation for $(2A+\phi)$. The general solution reads
\begin{equation}\label{Sol_alpha_beta_D}
\left\{
\begin{array}{lclc}
e^A & = & e^{A_0} c^{\frac{2}{D-3}} & ,\\[3mm]
e^B & = & e^{A_0} c^{\frac{2}{D-3}} & ,\\[3mm]
e^{\phi} & = & e^{-2A_0} \dfrac{(D-1)^2}{2(D-3)g^2L^2}c^{-\frac{2(D-1)}{D-3}}\left((D-3)+4s^2\right) & ,\\[3mm]
\lambda & = &\lambda_0+e^{2A_0}\sqrt{\dfrac{D+1}{D-3}} \dfrac{g^2L^2}{(D-1)^2} s\left((D+1)F_1\left(\frac{1}{2},2,\frac{D-7}{2(D-3)}, \left.\frac{3}{2}\right| \frac{-4}{D-3}s^2,s^2\right)\right. & \\
&  & \phantom{\lambda_0+e^{2A_0}\sqrt{\dfrac{D+1}{D-3}} \dfrac{g^2L^2}{(D-1)^2} s}\left.-(D-3)F_1\left(\frac{1}{2},\frac{D-7}{2(D-3)},1,\left.\frac{3}{2}\right|s^2,\frac{-4}{D-3}s^2\right)\right) & ,
\end{array}\right.
\end{equation}
where we introduced the shorthand notation 
\begin{equation}
\begin{array}{lclc}
c & \equiv & \cos\left(\frac{\sqrt{(D-1)(D-3)}}{2L}\left(y-y_0\right)\right) & , \\[2mm]
s & \equiv & \sin\left(\frac{\sqrt{(D-1)(D-3)}}{2L}\left(y-y_0\right)\right) & ,
\end{array}
\end{equation}
with $A_0, y_0, \lambda_0$ real integration constants. $F_1$ here denotes the Appell hypergeometric function (see appendix \ref{appendix} for further details). It may be worth noticing that the above family of solutions for generic $D$ is strictly defined only when $D> 3$, as it seems complex for $D=2$ and singular for $D=3$. However, the solution for $D=2$ makes perfect sense, upon denoting by $c$ \& $s$ $\cosh$ \& $i\sinh$, respectively. Almost everywhere $s$ appears through its square (so it is real), except for a factor $s$ in the expression for $\lambda$, but this gets compensated by $1/\sqrt{D-3}$ that picks up an $i$ as well for $D=2$. The $D=3$ case must be treated separately. By doing so, one finds
\begin{equation}\label{Sol_alpha_beta_3}
\left\{
\begin{array}{lclc}
e^A & = & e^{A_0} e^{-\frac{\left(y-{y_0}\right)^2}{L^2}} & ,\\[3mm]
e^B & = & e^{A_0} e^{-\frac{\left(y-{y_0}\right)^2}{L^2}} & ,\\[3mm]
e^{\phi} & = & e^{-2A_0} \dfrac{2}{g^2 L^2} \left(1+2 \dfrac{\left(y-{y_0}\right)^2}{L^2}\right) e^{\frac{\left(y-{y_0}\right)^2}{L^2}}& ,\\[3mm]
\lambda & = &\lambda_0+e^{2A_0} \dfrac{g^2 L^2}{2\sqrt{2}} \left( \dfrac{2L\left(y-{y_0}\right)}{L^2+2\left(y-{y_0}\right)^2} e^{-\frac{\left(y-{y_0}\right)^2}{L^2}} +\sqrt{\pi}\,\textrm{erf}\left(\frac{y-y_0}{L}\right)\right)& ,
\end{array}\right.
\end{equation}
where erf denotes the error function (see again appendix \ref{appendix} for further details).

\begin{figure}[!http]
\centering
\begin{tabular}{lcl}
 \includegraphics[scale=0.55]{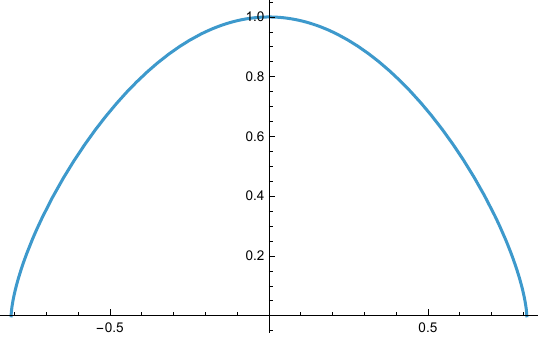} & \includegraphics[scale=0.55]{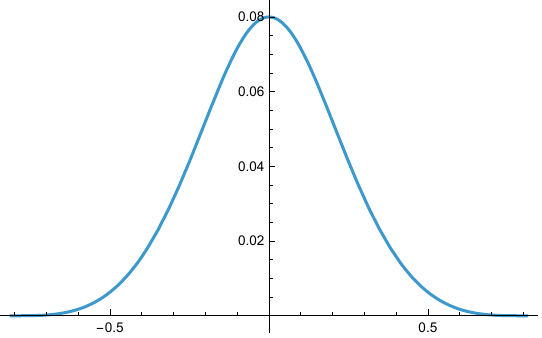} & \includegraphics[scale=0.55]{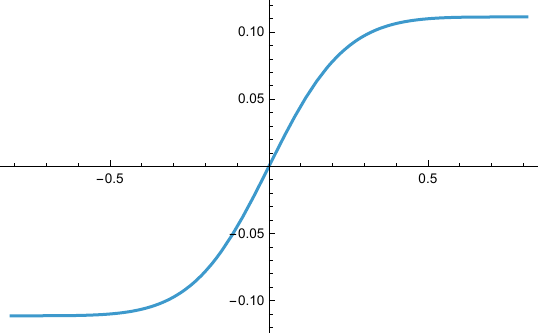}
\end{tabular}         
\caption{\it Plots of $e^A$ (\emph{Left}),  $e^{-\phi}$ (\emph{Center}), and $\lambda$ (\emph{Right}), for the explicit solution in the family \eqref{Sol_alpha_beta_D} with $D=6$, $A_0=y_0=\lambda_0=0$, $g=L=1$.}
\label{Sol_alphabeta}
\end{figure}

Now that the solutions are presented, we want to discuss their physical properties. Let us start from plotting the concrete behavior of the solution we just found. This can be done for a generic value of $D$, as it may be checked that its qualitative behavior does not crucially depend on $D$. The relevant plots for \emph{e.g.} $D=6$ may be found in figure \ref{Sol_alphabeta}.
These solutions have an $\mathrm{AdS}_D\times I$ topology, where the effective size of AdS shrinks at both ends of the interval $I$. However, at the same time, the dilaton diverges. The endpoints of $I$ are therefore singularities that should be studied more thoroughly. A first comment one can make is that such singuarities can be resolved in higher dimensions, at least for a special solution in this class, \emph{i.e.} the $D=7$ one. For this particular case, the solution gets significantly simplified, to yield
\begin{equation}
\left\{
\begin{array}{lclc}
e^A & = & e^B \ = \ e^{A_0}c^{1/2} & , \\[2mm]
e^{\phi} & =& e^{-2A_0} \dfrac{18}{g^2L^2} \dfrac{1+s^2}{c^3} & , \\[3mm]
\lambda & = &\lambda_0+ e^{2A_0}  \dfrac{\sqrt{2}g^2L^2}{9} \dfrac{s}{1+s^2} & ,
\end{array}\right.
\end{equation}
which exactly lifts to the non-supersymmetric $\mathrm{AdS}_7\times S^4$ vacuum of 11D supergravity, upon using the 11D uplift formulae given by \cite{Salam:1984ft}. The singularities at the endpoints of our interval get lifted to the poles of the $S^4$, in such a way that the 11D geometry be perfectly smooth.

In order to investigate the nature of the singularities for arbitrary $D$, we can have a look at the FP determined by the asymptotic values of $(X,Y,Z)$
\begin{equation}
(X,Y,Z) \ \overset{y \rightarrow \partial I}{\longrightarrow} \  \left(-\sqrt{\frac{D-1}{2D}}, \ 0,\ i\sqrt{\frac{D+1}{2D}}\right) \ ,
\nonumber
\end{equation}
which belongs to family IV. Therefore, we expect the singularities at the boundaries of $I$ to signal the presence of ETW branes. If we go back to the special $D=7$ case, these are nothing but KK monopoles filling $\mathrm{AdS}_7$, which are indeed resolved into 11D geometry.

\subsection{Type IIB-like scenarios: $m^2=0$, arbitrary $\beta$}
Let us now move to a setup with no potential, \emph{i.e.} $m^2=0$, but keeping the axion kinetic coupling $\beta$ arbitrary. By adopting the gauge where $B=DA-2\beta\phi$, the equation of motion of $\lambda$ simply boils down to $\lambda'= \textrm{const} \equiv Q$. This way, the equation of motion of $\phi$ gets decoupled from $A$ and may be directly integrated. By analogy to the previous case, we then decided to switch to a slightly different gauge, where $y$ is an angle. This yields
\begin{equation}\label{Sol_IIB_D}
\left\{
\begin{array}{lclc}
e^A & = & e^{A_0}\left(e^{A_1}\left(\dfrac{c}{s}\right)^{\beta_D}+e^{-A_1}\left(\dfrac{c}{s}\right)^{-\beta_D}\right)^{-\frac{1}{D-1}}  & , \\[2mm]
e^B & = & e^{A_0}\dfrac{QL}{\gamma_D}\dfrac{1}{sc}\left(e^{A_1}\left(\dfrac{c}{s}\right)^{\beta_D}+e^{-A_1}\left(\dfrac{c}{s}\right)^{-\beta_D}\right)^{-\frac{D}{D-1}} & , \\[5mm]
e^{\beta \phi} & = &  e^{-(D-1)A_0}\dfrac{QL}{\gamma_D}\dfrac{1}{sc} & , \\[3mm]
\lambda & = &  \lambda_0 - e^{(D-1)A_0}\dfrac{\gamma_D}{QL\beta}\left(1-2s^2\right) & ,
\end{array}\right.
\end{equation}
where we introduced the shorthand notation 
\begin{equation}
\begin{array}{lclc}
c & \equiv & \cos\left(\frac{Q\beta}{2}\left(y-y_0\right)\right) & , \\[2mm]
s & \equiv & \sin\left(\frac{Q\beta}{2}\left(y-y_0\right)\right) & ,
\end{array}
\end{equation}
with $\beta_D \equiv \frac{D-1}{\beta\gamma_D}$, and $A_0$, $A_1$, $y_0$ and $ \lambda_0$ are integration constants.

\begin{figure}[!http]
\centering
\begin{tabular}{lcl}
 \includegraphics[scale=0.55]{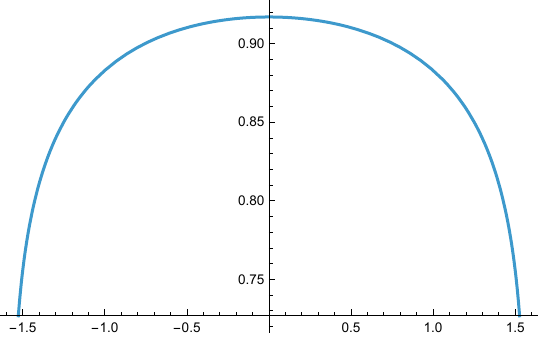} & \includegraphics[scale=0.55]{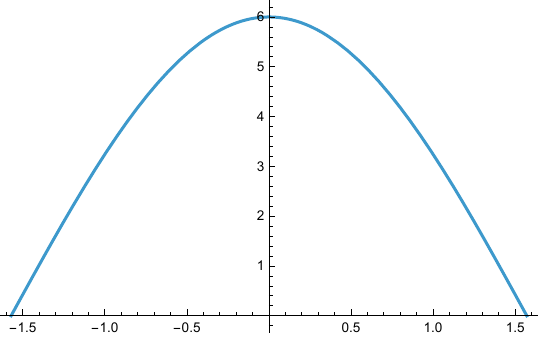} & \includegraphics[scale=0.55]{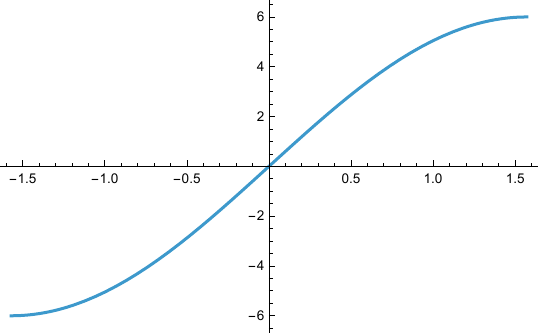}
\end{tabular}         
\caption{\it Plots of $e^A$ (\emph{Left}),  $e^{-\phi}$ (\emph{Center}), and $\lambda$ (\emph{Right}), for the explicit solution in the family \eqref{Sol_IIB_D} with $D=9$, $A_0=A_1=\lambda_0=0$, $y_0=-\frac{\pi}{2}$ and $Q=L=1$.}
\label{Sol_betaB}
\end{figure}

As in previous cases, the qualitative behavior of these solutions does not crucially change with $D$. For an explicit demonstration, we decided to plot the most relevant solution in this class, \emph{i.e.} $D=9$, $\beta=1$, corresponding to a type IIB $\mathrm{AdS}_9\times I$ solution. We refer to our companion paper \cite{Dibitetto:2026oor} for more details and discussions about this special case. The corresponding plots may be found in figure \ref{Sol_betaB}.

These solutions present again an $\mathrm{AdS}_D\times I$ topology, with a compact radial slicing. When Wick-rotated to Euclidean signature, they have the typical features of Euclidean wormholes. Indeed, by directly evaluating the on-shell Euclidean action, we find that it is finite and reads
\begin{equation}
S_{\textrm{E}}|_{\textrm{on-shell}} \ = \ e^{(D-1)A_0}D L^{D-1}\dfrac{\mathrm{Vol}\left(\mathbb{H}^D\right)}{\kappa_{D+1}^2} \ ,
\end{equation}
where $\mathrm{Vol}\left(\mathbb{H}^D\right)$ denotes the regularized volume of $D$ dimensional hyperbolic space, and the only non-zero contribution to the on-shell action comes from the GHY boundary term evaluated at the endpoints of our interval $I$.
To further evidence for the finiteness of the Euclidean action, we notice that the FP that describe the asymptotic regime around the singularities at the endpoints of the flow are
\begin{equation}
(X,Y,Z) \ \overset{y \rightarrow \partial I}{\longrightarrow} \  \left(-1, \ 0,\ 0\right) \ ,
\nonumber
\end{equation}
which belongs to family I, as it should, since it is the only stable FP in the corresponding region of the phase diagram. Family I is exactly the one in which the singularity gets resolved in one higher dimension into a smooth geometry.

\subsection{Massive IIA-like scenarios: no axion and $m^2>0$: some special cases}
The third class of solutions we want to discuss is the one where the axion vanishes identically $\lambda\equiv 0$, the only ingredient supporting the flow being the $m^2$ driven potential for $\phi$. Massive type IIA supergravity is in principle included here, by choosing $D=9$, $m^2>0$ and $\alpha=\frac{5}{4}$. Unfortunately though, within this class of models, we were only able to analytically integrate the field equations for special values of $\alpha$, and the mIIA case was not among those. After making some general observations about the equations in this case, we'll present the analytic classes we could find. 

First of all, when the axion vanishes identically its equation of motion is trivially satisfied, while the $\phi$ equation may be written as
\begin{equation}
\left(e^{DA-B}\phi'\right)' \ = \ m^2 e^{DA+B+2\alpha\phi} \ \overset{\textrm{gauge fix}}{=} m^2 \ ,
\end{equation}
which holds in the $B=-DA-2\alpha\phi$ gauge. With this choice, one may express $A$, $\phi$ and hence $B$ all in terms of a new function $G(y)$ and its derivative
\begin{equation}
\begin{array}{lccclc}
e^{-2DA} \ = \ y^{-1}G' \ , & & & e^{2\alpha\phi} \ = \ 4m^2\alpha^2 G & .
\end{array}
\end{equation}
This way, the field equations and the Hamiltonian constraint are implied by the following non-linear ODE for $G$
\begin{align}
G^{(3)}+\frac{G'-y G''}{2yG'}\left(\frac{3D+1}{D} G'' +\frac{D-1}{D}\frac{G'}{y}\right) + \phantom{\left(\frac{3D+1}{D} G'' +\frac{D-1}{D}\frac{G'}{y}\right)} \notag\\
+\frac{G' }{G}\left(G''-\frac{ \left(2 \alpha ^2
   (D-1)+1\right) G'}{2\alpha ^2 (D-1) y}\right)-\frac{G'^3}{4\alpha ^2 G^2} \,=\, 0 \ ,
\end{align}
Unfortunately, the above ODE does not seem to enjoy special properties for generic $\alpha$. Nevertheless, in the following we'll present three different cases where analytic solutions may be found.
\begin{itemize}
\item $\alpha \,=\, \frac{D+1}{2\gamma_D}$: In this case the optimal gauge is $B=A$. The solution reads
\end{itemize}
\begin{equation}\label{Sol_IIA1_D}
\left\{
\begin{array}{rclc}
e^{A} & = &\left(\mu sc-\dfrac{e^{2\psi_0}m^2L^2}{D(D-1)} c^2 {}_2F_1\left(-\frac{1}{2},\frac{D+1}{D-1},\frac{1}{2}\left|-\frac{s^2}{c^2}\right.\right)\right)^{\frac{1}{D-1}}  & , \\[4mm]
e^{B} & = &\left(\mu sc-\dfrac{e^{2\psi_0}m^2L^2}{D(D-1)} c^2 {}_2F_1\left(-\frac{1}{2},\frac{D+1}{D-1},\frac{1}{2}\left|-\frac{s^2}{c^2}\right.\right)\right)^{\frac{1}{D-1}} & , \\[5mm]
e^{\alpha \phi} & = &  e^{\psi_0}c^{\frac{D+1}{D-1}}\left(\mu sc-\dfrac{e^{2\psi_0}m^2L^2}{D(D-1)} c^2 {}_2F_1\left(-\frac{1}{2},\frac{D+1}{D-1},\frac{1}{2}\left|-\frac{s^2}{c^2}\right.\right)\right)^{-\frac{D+1}{2(D-1)}} & , \\[3mm]
\lambda & = & 0 & ,
\end{array}\right.
\end{equation}
where we introduced the shorthand notation 
\begin{equation}
\begin{array}{lclc}
c & \equiv & \cos\left(\frac{D-1}{2L}\left(y-y_0\right)\right) & , \\[2mm]
s & \equiv & \sin\left(\frac{D-1}{2L}\left(y-y_0\right)\right) & ,
\end{array}
\end{equation}
with $\mu$, $\psi_0$ and $y_0$ are integration constants. ${}_2F_1$ here denotes the Gauss hypergeometric function (see appendix \ref{appendix} for further details).
\begin{itemize}
\item $\alpha \,=\, \frac{D-2}{\gamma_D}$: In this case the optimal gauge is $B=-\alpha \phi$. The solution reads
\end{itemize}
\begin{equation}\label{Sol_IIA2_D}
\left\{
\begin{array}{rclc}
e^{A} & = &\left(\mu sc^{D-1}-\dfrac{D e^{2\psi_0}}{m^2L^2}\dfrac{ c^{2(D-1)}}{s^2} {}_2F_1\left(1,\frac{3}{2},\frac{D+1}{2}\left|-\frac{c^2}{s^2}\right.\right)\right)^{\frac{1}{2(D-1)}}  & , \\[4mm]
e^{B} & = &e^{\psi_0}c^{D-2}\left(\mu sc^{D-1}-\dfrac{D e^{2\psi_0}}{m^2L^2}\dfrac{ c^{2(D-1)}}{s^2} {}_2F_1\left(1,\frac{3}{2},\frac{D+1}{2}\left|-\frac{c^2}{s^2}\right.\right)\right)^{-\frac{D-2}{2(D-1)}}  & , \\[5mm]
e^{\alpha \phi} & = &  e^{-\psi_0}c^{-(D-2)}\left(\mu sc^{D-1}-\dfrac{D e^{2\psi_0}}{m^2L^2}\dfrac{ c^{2(D-1)}}{s^2} {}_2F_1\left(1,\frac{3}{2},\frac{D+1}{2}\left|-\frac{c^2}{s^2}\right.\right)\right)^{\frac{D-2}{2(D-1)}} & , \\[3mm]
\lambda & = & 0 & ,
\end{array}\right.
\end{equation}
where we introduced the shorthand notation 
\begin{equation}
\begin{array}{lclc}
c & \equiv & \cos\left(\frac{m}{\sqrt{D}}\left(y-y_0\right)\right) & , \\[2mm]
s & \equiv & \sin\left(\frac{m}{\sqrt{D}}\left(y-y_0\right)\right) & ,
\end{array}
\end{equation}
with $\mu$, $\psi_0$ and $y_0$ are integration constants. ${}_2F_1$ here denotes the Gauss hypergeometric function (see appendix \ref{appendix} for further details).
\begin{itemize}
\item $\alpha \,=\, \frac{1}{\gamma_D}$: In this case the optimal gauge is $B=-(D-2)A$. The solution reads
\end{itemize}
\begin{equation}\label{Sol_IIA3_D}
\left\{
\begin{array}{rclc}
e^{A} & = &\left(\mu r-\dfrac{(D-1)^2}{L^2}r^2-\dfrac{(D-1)^2e^{2\psi_0}m^2}{D(D+1)}r^{\frac{2D}{D-1}}\right)^{\frac{1}{2(D-1)}}  & , \\[4mm]
e^{B} & = &\left(\mu r-\dfrac{(D-1)^2}{L^2}r^2-\dfrac{(D-1)^2e^{2\psi_0}m^2}{D(D+1)}r^{\frac{2D}{D-1}}\right)^{-\frac{D-2}{2(D-1)}} & , \\[5mm]
e^{\alpha \phi} & = &  e^{\psi_0}r^{\frac{1}{D-1}}\left(\mu r-\dfrac{(D-1)^2}{L^2}r^2-\dfrac{(D-1)^2e^{2\psi_0}m^2}{D(D+1)}r^{\frac{2D}{D-1}}\right)^{-\frac{1}{2(D-1)}} & , \\[3mm]
\lambda & = & 0 & ,
\end{array}\right.
\end{equation}
where we introduced the shorthand notation $r \, \equiv \,\left(y-y_0\right)$
with $\mu$, $\psi_0$ and $y_0$ are integration constants. It may be worth mentioning that this solution may be written in a similar fashion as the previous two by observing that $(y-y_0)^{\nu}$ is a degenerate hypergeometric function. In particular
\begin{equation}
(y-y_0)^{\frac{2D}{D-1}} \,=\, y^{\frac{2D}{D-1}}\,  {}_2F_1\left(-\frac{2D}{D-1},b,b\left|\frac{y_0}{y}\right.\right) \ ,
\end{equation}
for any arbitrary $b$.
In all three cases the asymptotic behavior is captured by the physics of FP in family I
\begin{equation}
(X,Y,Z) \ \overset{y \rightarrow \partial I}{\longrightarrow} \  \left\{
\begin{array}{llc}
 \left(-1, \ 0,\ 0\right) \ , & y\,\in\, \partial I^- & , \\
\left(+1, \ 0,\ 0\right) \ , & y\,\in\, \partial I^+ & .
\end{array}\right.
\nonumber
\end{equation}
This is to be expected, since all the corresponding  three values of $\alpha$ lie within region 'B' in the phase diagram depicted in figure \ref{fig:Phase_Diagram}. Because this FP represents a singularity that is typically resolved in one dimension higher, we expect these flows to have a finite on-shell Euclidean action, despite the presence of a non-zero potential. This feature complicates things in general, as it may not be guaranteed that the bulk contribution vanishes.

\begin{figure}[!http]
\centering
\begin{tabular}{lcl}
 \includegraphics[scale=0.55]{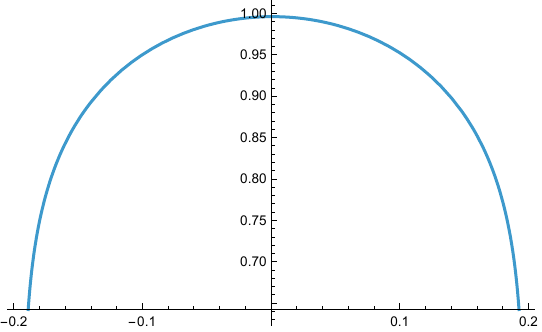} & \includegraphics[scale=0.55]{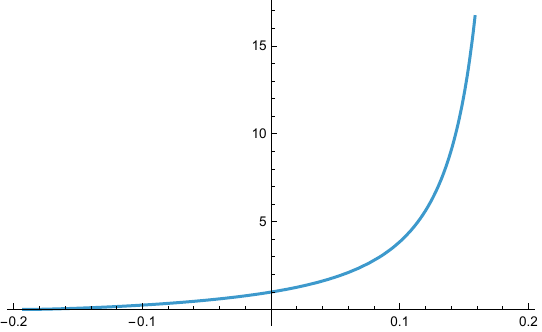} & \includegraphics[scale=0.55]{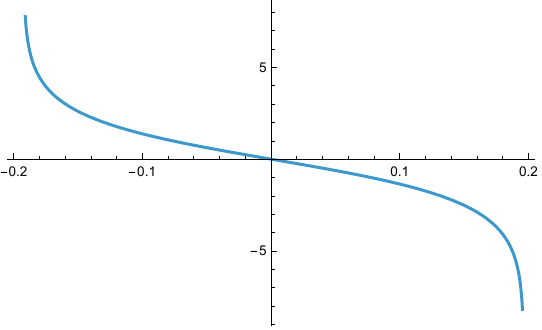}
\end{tabular}         
\caption{\it Plots of $e^A$ (\emph{Left}),  $e^{-\phi}$ (\emph{Center}), and $\phi$ (\emph{Right}), for the explicit solution in the family \eqref{Sol_IIA1_D} with $D=9$, $\psi_0=\alpha=\frac{D+1}{2\gamma_D}=\frac{5}{12}$, $y_0=-\frac{\pi L}{2(D-1)}=-\frac{\pi}{16}$, $\mu=1$ and $m=L=1$. The qualitative behavior is that of a compact $\mathrm{AdS}_9\times I$ slicing, the dilaton singularities at the endpoints of I being resolved into one dimension higher.}
\label{Sol_m1}
\end{figure}

As an example, we display  in figure \ref{Sol_m1} the plots of a solution within the first class, \emph{i.e.} with $\alpha=\frac{D+1}{2\gamma_D}$, and choose $D=9$. 

\subsection{Massive IIA-like scenarios: numerical flows for generic $\alpha$'s}
As anticipated earlier, we want to conclude by discussing the salient features of numerical flows with vanishing axion for different values of $\alpha$ than the ones for which we could find analytic solutions. For our three analytic families we have seen that they all end up in a FP of type I. We expect this to be no longer the case once we cross the critical value defined in \eqref{alpha_crit}. For larger $\alpha$'s than the critical value, we enter region 'A' in the phase diagram, where it's FP in family IV that become stable.

\begin{figure}[!http]
\centering
\begin{tabular}{ll}
 \includegraphics[scale=0.7]{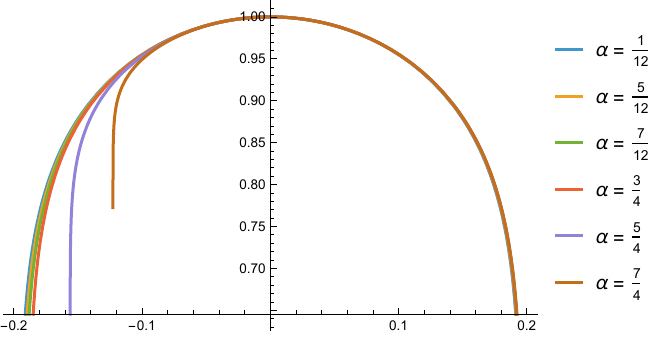} &  \includegraphics[scale=0.7]{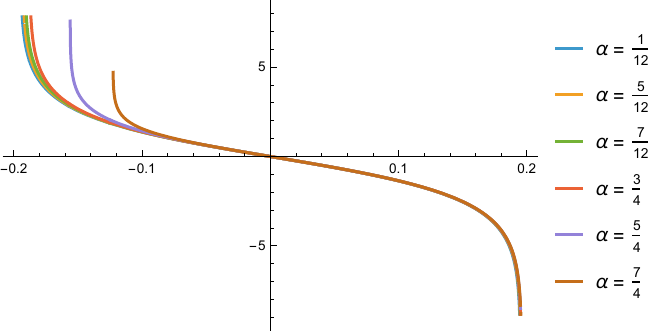}
\end{tabular}         
\caption{\it Plots of $e^A$ (\emph{Left}) and $\phi$ (\emph{Right}) for different numerical solutions with vanishing axion and $m^2>0$. Different colors represent different values of $\alpha$. The first four curves are almost identical, as they all interpolate between I$_-$ \& I$_+$. In the last two curves the I$_-$ FP is replaced by FP $IV$, which is to be interpreted as an ETW brane singularity.}
\label{fig:Num_alpha}
\end{figure}

For the sake of comparison, we have performed a numerical integration of the field equations for different values of $\alpha$, but sticking with the \emph{same gauge choice} $B=A$. We also fix $D=9$, since it includes the massive IIA case of interest, and also fix $m=L=1$, as a choice of units. For $D=9$, the analytic solutions presented in the previous section correspond to $\alpha\,=\, \frac{1}{12}; \ \frac{5}{12}; \ \frac{7}{12}$, all smaller than the critical value which is $\frac{3}{4}$. The massive type IIA theory is obtained for $\alpha=\frac{5}{4}$. In figure \ref{fig:Num_alpha} we display the polts for $e^A$ \& $\phi$ for several numerical solutions with different choices of $\alpha$. It may be seen that, moving backwards from $\partial I^+$ towards the left, it is only when approaching the singularity in  $\partial I^-$ that the field profiles start differing significantly from one another.
\begin{figure}[!http]
	\centering
	\includegraphics[width=0.75\textwidth]{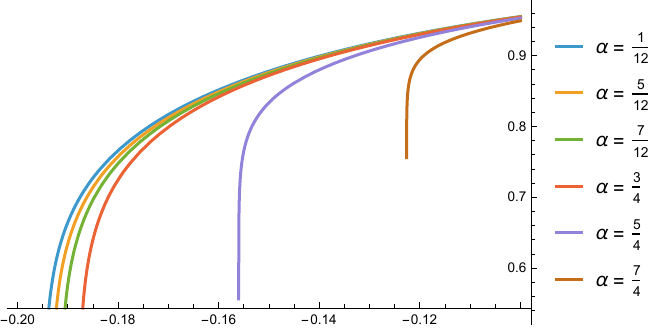}
	\caption{\it  A zoom of the plot for  $e^A$ onto the region where all solutions significantly move apart from one another. Different colors still indicate different values of $\alpha$.}
	\label{fig:Num_alpha_Zoomed}
\end{figure}

\begin{figure}[!http]
	\centering
	\includegraphics[width=0.75\textwidth]{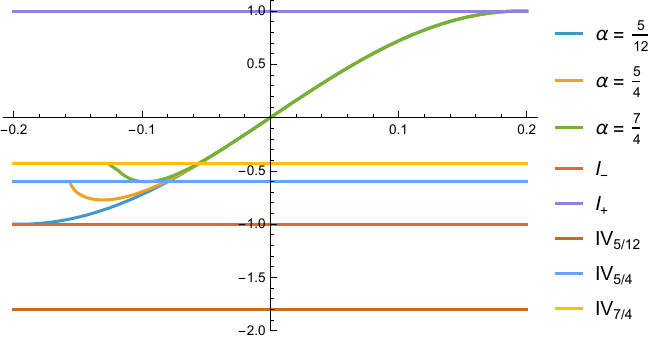}
	\caption{\it  The evolution of $X^{-1}$ along our numerical flows for different $\alpha$'s. It may be seen that for subcritical models, FP IV corresponds to values of $X^{-1}$ below $-1$, which are never reached. In critical models FP IV lies exactly at $-1$, thus coinciding with I$_-$. In overcritical models  FP IV gets preferred to I$_-$ because it has a higher value of $X^{-1}$. Note that the left endpoints of the different flows do not coincide and moreover, $X^{-1}$ does not need to approach the FP values with a flat slope, as this will only be the case once expressed as a function of $A$ rather than $y$.	}
	\label{fig:FP_alpha}
\end{figure}

All the analytic solutions presented in this paper were essentially \emph{sysmmetric}, in the sense that the FP at the endpoints of the interval $I$ were the same, except for the three analytic classes of massive IIA-like flows, which still interpolate between FP of family I, but with different signs $\sigma\in\mathbb{Z}_2$ at the two ends. This sign flip occured because of the $y$-odd profile of the dilaton $\phi$. Now, for $\alpha$'s above the critical value, we have \emph{truly asymmetric} flows in a deeper sense, as they interpolate between FP of different families. Indeed, one finds 
\begin{equation}
\begin{array}{lcr}
(X,Y,Z) \ \overset{y \rightarrow \partial I}{\longrightarrow} \  \left\{
\begin{array}{llc}
 \left(-1, \ 0,\ 0\right) \ , & y\,\in\, \partial I^- & , \\
\left(+1, \ 0,\ 0\right) \ , & y\,\in\, \partial I^+ & ,
\end{array}\right.  & &
(\text{for }\alpha \,<\, \alpha^{(D)}_* )
\end{array}
\nonumber
\end{equation}
while, on the other hand,
\begin{equation}
\begin{array}{lcr}
(X,Y,Z) \ \overset{y \rightarrow \partial I}{\longrightarrow} \  \left\{
\begin{array}{llc}
 \left(-\frac{\gamma_D\alpha}{D}, \ 0,\ -\sqrt{\frac{\gamma^2_D\alpha^2}{D^2}-1}\right) \ , & y\,\in\, \partial I^- & , \\
\left(+1, \ 0,\ 0\right) \ , & y\,\in\, \partial I^+ & .
\end{array}\right. & & (\text{for }\alpha \,>\, \alpha^{(D)}_*)
\end{array}
\nonumber
\end{equation}
It is worth noticing that exactly at the critical value $\alpha^{(D)}_*$ the two expressions coincide (FP I$_-$ \& FP IV$_-$). In figure \ref{fig:FP_alpha} we plot the inverse of $X$ (as defined in \eqref{XYZ_def}) along the flow, for different values of $\alpha$, including the massive IIA one. Since the corresponding ETW singularity is a generalization of a D8/O8 singularity, we expect these flows to have a divergent Euclidean action, as also noted in our companion paper \cite{Dibitetto:2026oor}.

\section{Final Comments}
We would like to conclude our analysis by raising a few related issues to which we only have partial answers at the moment and we do plan to continue addressing in the future.
\begin{itemize}
\item \textbf{Finiteness of the Euclidean action.} We stressed that some of these solutions may be relevant in the context of non-spersymmetric holography. In this respect it is crucial to assess whether or not the onshell Euclidean action is finite. This question is tightly related to the compactness of the solution. Generically, it is not a straightforward thing to check, at least when $m^2\neq 0$, since in that case one cannot simply conclude that the bulk contribution vanishes. Our expectations though are such that all flows connecting FP from family I, II and III should have a finite Euclidean action, as they all get resolved into smooth higher-dimensional geometries. When the solution asymptotes to a FP of type IV, the issue is much more delicate since the corresponding flow ends up in an ETW brane singularity. Depending on the nature of the associated source, the on-shell action may be finite or not. For example, we commented on its finiteness when the ETW sources are KK monopoles, while we argued for its divergence in the massive IIA case. Finally, even for those cases where it is divergent, one may still wonder whether it is possible to come up with a regularization scheme that unambiguously extracts its finite part.
\item \textbf{Analytic continuations $L \ \rightarrow \ iL$ and dS.} In the general field equations \eqref{D_EOMs}, this operation simply amounts to flipping the sign of the $ \frac{1}{L^2}$ term associated with the slice curvature. Effectively, solving the equations in this case means to look for $\mathrm{dS}_D\times I$ solutions. By taking an explicit look at our analytic solutions, we may simply try to see whether the fields remain real after a Wick rotation of $L$, possibly by absorbing complex quantities within the integration constants and by changing the range of the radial coordinate $y$. In the family of solutions in  \eqref{Sol_alpha_beta_D}, this seems impossible, as it automatically produces an \emph{immaginary axion}. A similar story seems to apply to \eqref{Sol_IIB_D}, where it is instead the dilaton that gets a factor of $i$. A very different situation occurs in mIIA-like scenarios considered in the third part. Indeed, in \eqref{Sol_IIA1_D}, we could rewrite everything in terms of the same $c$ \& $s$, which would now represent $\cosh$ \& $i\sinh$, respectively. The field profiles then only depend on $L^2$ and $s^2$, except for the term proportional to the integration constant $\mu$. So it is enough to choose imaginary values of $\mu$ to get a real solution.
For \eqref{Sol_IIA2_D} and \eqref{Sol_IIA3_D} the situation is even simpler because we only have $L^2$ dependence and a sign flip of those terms is harmless for the reality of the solutions. As a demonstration, we plot the corresponding dS solution in figure \ref{Sol_m4}.
\begin{figure}[!http]
\centering
\begin{tabular}{lcl}
 \includegraphics[scale=0.52]{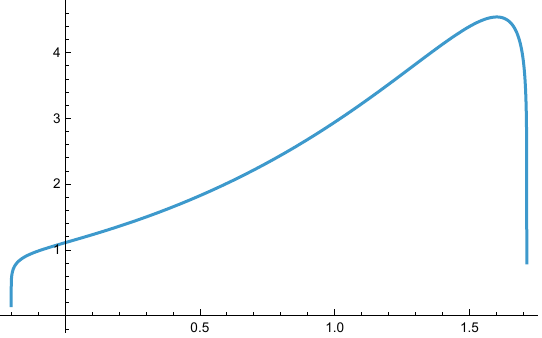} & \includegraphics[scale=0.52]{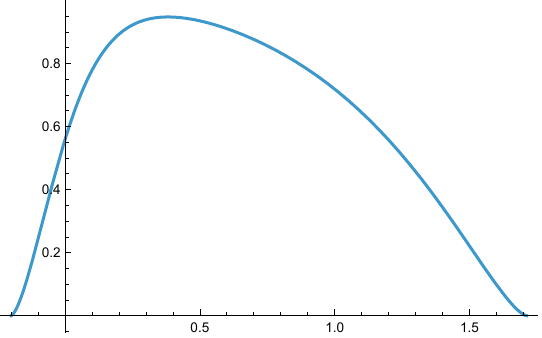} & \includegraphics[scale=0.52]{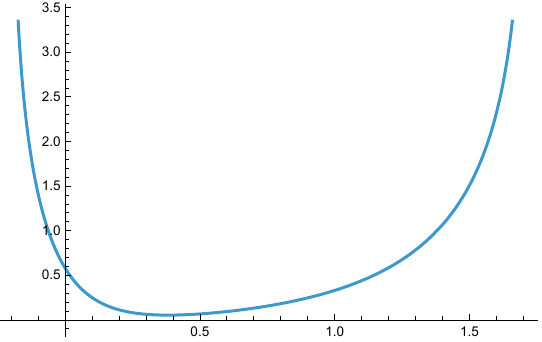}
\end{tabular}         
\caption{\it Plots of $e^A$ (\emph{Left}),  $e^{-\phi}$ (\emph{Center}), and $\phi$ (\emph{Right}), for the explicit $\mathrm{dS}_9\times I$ solution obtained from family \eqref{Sol_IIA1_D} by sending $L\rightarrow iL$ and $\mu\rightarrow i\mu$. In the plots we chose $D=9$, $\psi_0=\alpha=\frac{D+1}{2\gamma_D}=\frac{5}{12}$, $y_0=-\frac{\pi L}{2(D-1)}=-\frac{\pi}{16}$, $\mu=1$ and $m=L=1$. }
\label{Sol_m4}
\end{figure}
Solutions of this type seem to be compact, but evade the no-go theorem of \cite{Maldacena:2000mw} due to the presence of an $m^2$ potential.
\item \textbf{Relation to Euclidean wormholes.} This issue is much more speculative. Many of the solutions discussed here, certainly those where the on-shell Euclidean action is arguably finite, exhibit the typical salient features of Euclidean wormholes. These are compact slicings of the $(D+1)$ dimensional metric with maximally symmetric co-dimension one slices, with two distinct boundaries. The extra feature that our solutions do not have though, is that wormholes usually link two different vacuum solutions. Nevertheless, these novel solutions discussed in our work seem to pose similar problems for the definition of quantum gravity path integrals. We plan to elaborate more on this aspect in the future.
\item  \textbf{Analytic continuations $m \ \rightarrow \ im$ and string/M theory embeddings.} Throughout the paper, $m^2$ is allowed to have either signs. In the three different groups of analytic solutions that we presented, the first one had $m^2<0$, the second one $m^2=0$ and the third one $m^2>0$. The sign of the potential plays a crucial role in understanding the possible higher dimensional picture of a given model. Details concerning this aspect are discussed in appendix \ref{appendix_KK}. Similarly to what discussed for Wick rotations of $L$, the first family with $\alpha=\frac{1}{2}$ and $\beta=1$ strictly requires $m^2=-g^2<0$ and does not allow this analytic continuation. Type IIB-like scenarios behave trivially since that have no potential to start with. Finally, mIIA-like solutions can all be mapped into solutions with the opposite sign of $m^2$.
As for stringy embeddings of our models, we cannot help noticing that all $(\alpha,\beta)$ models that admit a higher dimensional description in the sense explained in appendix \ref{appendix_KK} come from Einstein gravity + a $d$-form in ten or eleven dimensions. This may of course be a sector of the full field content of type II supergravities or 11D supergravity. It would be interesting to explicitly construct the 10D/11D lift of our analytic solutions to see whether these yield already known or novel non-supersymmetric AdS string backgrounds.
\end{itemize}

\section*{Acknowledgements}

The work of GD is partly supported by an INFN fellowship under the “Iniziativa Specifica” ST\&FI. NP thanks the members of the hep-th group at the University of Turin for their kind hospitality while parts of this work were carried out.

\appendix

\section{Dimensional reductions and axio-dilaton gravity}
\label{appendix_KK}
Let us analyze the possible higher dimensional origin of the models considered in this work. It is reasonable to imagine that a standard way of obtaining the $(D+1)$ dimensional action \eqref{action} is to start from Einstein gravity in $(D+d+1)$, coupled to a $d$-form gauge potential
\begin{equation}\label{action_Dd1}
S_{D+d+1} \ = \ \frac{1}{2\kappa_{D+d+1}^2}\int d^{D+d+1}x \sqrt{-g_{D+d+1}}\,\left(\mathcal{R}_{D+d+1}-\frac{1}{2(d+1)!}\left|F_{(d+1)}\right|^2\right) \ ,
\end{equation}
where $\kappa_{D+d+1} $ is the gravitational coupling in $(D+d+1)$ dimensions, while $F_{(d+1)} \,\equiv\, dC_{(d)}$. Reducing the above action on a compact  $d$ dimensional Einstein space $\mathcal{M}_d$ will generically produce a runaway potential in the KK scalar $\phi$, coming from the internal curvature, and an axion coming from the gauge potential wrapping  $\mathcal{M}_d$. Our reduction \emph{Ansatz} reads
\begin{equation}\label{KK_Ansatz}
\begin{array}{lclc}
ds_{D+d+1}^2 & = & e^{2\xi\phi} ds_{D+1}^2 \, + \, e^{2\zeta\phi}ds_{\mathcal{M}_d}^2 & , \\
C_{(d)} & = & \lambda \, \mathrm{vol}_{\mathcal{M}_d} & ,
\end{array}
\end{equation}
where $\xi$ \& $\zeta$ are suitable constants, while $\phi$ \& $\lambda$ are assumed to be lower dimensional (pseudo)scalar degrees of freedom.  Imposing that our reduction procedure yield $(D+1)$ dimensional gravity in the \emph{Einstein frame} coupled to a \emph{canonically normalized} $\phi$ completely fixes the constants $\xi$ \& $\zeta$ (up to a sign, which is fixed by assuming $\xi>0$) through
\begin{equation}\label{xizeta}
\begin{array}{lccclc}
\xi \ = \ \sqrt{\dfrac{d}{2(D-1)(D+d-1)}} & & \text{and} & & \zeta \ = \ - \sqrt{\dfrac{D-1}{2d(D+d-1)}} & .
\end{array}
\end{equation}

If we now specify the Ricci scalar of $\mathcal{M}_d$ as $\mathcal{R}_d\,=\,\frac{d(d-1)}{R^2}$, where $R$ is the radius, we exactly reproduce an action of the form of \eqref{action}, upon performing the following identifications
\begin{equation}\label{KK_dictionary}
\left\{
\begin{array}{lclc}
m^2 & = & -\dfrac{d(d-1)}{R^2} & , \\[3mm]
\alpha & = & \sqrt{\dfrac{D+d-1}{2d(D-1)}} & , \\[3mm]
\beta & = & \sqrt{\dfrac{d(D-1)}{2(D+d-1)}} & ,
\end{array}\right.
\end{equation}
where $m^2$ might have either signs, depending on the sign of the curvature of $\mathcal{M}_d$ (in this notation, we would need to send $R\rightarrow i R$ to flip the sign of the internal curvature).
This implies that only those models obeying the above constraints for a suitable integer $d$ admit a higher dimensional origin of this form. In particular, $\alpha^2$ \& $\beta^2$ must be rational numbers, and $\alpha\beta \,\overset{!}{=} \, \frac{1}{2}$. Note that this is exactly the case for our first family of analytic solutions, where $\alpha =\frac{1}{2}$ and $\beta=1$ .  Indeed, by exploring the uplift conditions for the $ (\frac{1}{2}, 1)$ models, we find
\begin{equation}
d \ \overset{!}{=} \ \frac{2(D-1)}{D-3}\ \overset{!}{\in} \ \mathbb{N} \ ,
\end{equation}
which is \emph{e.g.} solved for $(D,d) \ = \ (7,3) ; \ (5,4) ; \ (4,6)$. 

In models with identically vanishing axion like the ones considered in the last part of our anlaysis, we only need to satsify the first two constraints in \eqref{KK_dictionary}. The former one will simply tell us whether we need positive or negative internal curvature, depending on the sign of $m^2$. The latter one, may be re-expressed as a constraint for $d$ given $\alpha$
\begin{equation}
d \ = \ \frac{D-1}{2(D-1)\alpha^2-1} \ = \ \left\{
\begin{array}{lclc}
\frac{4D}{D-1} & , & \alpha\,=\,\frac{D+1}{2\gamma_D} & , \\[3mm]
\frac{D}{D-4} & , & \alpha\,=\,\frac{D-2}{\gamma_D} & , \\[3mm]
-D & , & \alpha\,=\,\frac{1}{\gamma_D} & .
\end{array}
\right.
\end{equation}
For our three analytic families of solutions, this implies no higher dimensional origin of this type for the third one. As for the remaining two, there are good values of $D$ for which this uplifting procedure works out. When $\alpha\,=\,\frac{D+1}{2\gamma_D}$, we need to pick out those values of $D$ for which $\frac{4D}{D-1}$ is a natural integer. This yields $(D,d) \ = \ (2,8) ; \ (3,6) ; \ (5,5)$.
When $\alpha\,=\,\frac{D-2}{\gamma_D}$, we need $D$'s for which $D-4$ divides $D$, selecting special cases such as \emph{e.g.}  $(D,d) \ = \ (5,5) ; \ (6,3) ; \ (8,2)$. In all cases where this geometrizing procedure happens to succeed, $(D+d+1)$ turns out to be \emph{either ten or eleven}.

\section{Hypergeometric functions and their properties}
\label{appendix}
Hypergeometric functions are special functions defined on the Riemann sphere $\hat{\mathbb{C}}\,\equiv\, \mathbb{C}\cup \{\infty\}$. Within the unit ball (\emph{i.e.} $|z|<1)$, they are defined through the so-called hypergeometric series
\begin{equation}
{}_2F_{1}(a,b,c|z) \, \equiv \, \sum\limits_{k=0}^{+\infty}\frac{(a)_k(b)_k}{(c)_k} \, z^k \, =\, 1+\frac{ab}{c}\,z+\frac{a(a+1)b(b+1)}{c(c+1)}\,\frac{z^2}{2}+\dots\ ,
\end{equation}
where the so-called Pochhammer symbol $(q)_k$ is defined through
\begin{equation}
(q)_k \, \equiv \, \left\{
\begin{array}{lccc}
1 \ , & & k=0 & , \\
q(q+1) \dots (q+k-1) \ , & &  k\geq 1 & .
\end{array}
\right.
\end{equation}
Hypergeometric functions are characterized as solutions of a second order Fuchsian ODE with three regular singularities, respectively placed at $z=0$; $1$ and $\infty$
\begin{equation}
z(1-z)\,F''+\big(c-(a+b+1)z\big)\,F'-ab\,F \ = \ 0 \ .
\end{equation}
Moreover, any other second order Funchsian ODE with three regular singularities may be brought into the above form upon performing a suitable change of variables. Because the space of solutions is everywhere two-dimensional, there exists a set of identities \emph{a.k.a.} connection formulae, that allow one to analytically continue the functions outside of the unit ball. These connection formulae are
\begin{equation}
\begin{array}{lclc}
{}_2F_{1}(a,b,c|1-z) &=& \dfrac{\Gamma(c)\Gamma(c-a-b)}{\Gamma(c-a)\Gamma(c-b)}{}_2F_{1}(a,b,c-a+b-c+1|z) +\\[3mm]
   & + &   \dfrac{\Gamma(c)\Gamma(a+b-c)}{\Gamma(a)\Gamma(b)}z^{c-a-b}{}_2F_{1}(c-a,c-b,c-a-b+1|z) &  ,
\end{array}
\end{equation}
and
\begin{equation}
\begin{array}{lclc}
{}_2F_{1}(a,b,c|z^{-1}) &=& \dfrac{\Gamma(c)\Gamma(b-a)}{\Gamma(b)\Gamma(c-a)}(-z)^a{}_2F_{1}(a,a-c+1,a-b+1|z) +\\[3mm]
   & + &   \dfrac{\Gamma(c)\Gamma(a-b)}{\Gamma(a)\Gamma(c-b)}(-z)^b{}_2F_{1}(b,b-c+1,b-a+1|z) &  , \\
\end{array}
\end{equation}
where $\Gamma(w)$ denotes Euler's gamma function.
The above formulae turn out to be extremely useful in order to study the behavior of hypergeometric functions around $z=1$ and $z=\infty$, respectively. 

Hypergeometric functions also obey the following differentiation formula
\begin{equation}
\frac{d^n}{dz^n}{}_2F_{1}(a,b,c|z) \,=\, \frac{(a)_n(b)_n}{(c)_n}\,{}_2F_{1}(a+n,b+n,c+n|z) \ .
\end{equation}

\subsection*{Degenerate cases}
\begin{itemize}
\item ${}_2F_{1}(a,b,b|z) \,=\, (1-z)^{-a}$, for any $b$. If we choose $a=1$ in particular, it reduces to the sum of the geometric series, whence the name of hypergeometric.
\item Chebyshev polynomials: ${}_2F_{1}\left(a,-a,\frac{1}{2}|\sin^2\left(\frac{\theta}{2}\right)\right) \,=\,\cos(a\theta)$.
\item Confluent limit: ${}_1F_{1}(a,c|z) \,=\, \lim\limits_{b \rightarrow \infty}{}_2F_{1}(a,b,c|b^{-1}z)$, \emph{a.k.a.} Kummer's functions.
\end{itemize}
The \emph{error function} erf appearing in \eqref{Sol_alpha_beta_3} may then understood as a degnerate confluent hypergeometric function by virtue of
\begin{equation}
\mathrm{erf}(z) \, \equiv \, \frac{2}{\sqrt{\pi}}\int\limits_{0}^{z}e^{-t^2}dt \, = \, \frac{2z}{\sqrt{\pi}}{}_1F_{1}\left(\frac{1}{2},\left.\frac{3}{2}\right|-z^2\right) \ .
\end{equation}

Finally, the \emph{Appell function} appearing in \eqref{Sol_alpha_beta_D} is defined through a hypergeometric series in two complex variables
\begin{equation}
F_{1}(a,b_1,b_2,c|z,w) \, \equiv \, \sum\limits_{k,\ell=0}^{+\infty}\frac{(a)_{k+\ell}(b_1)_k(b_2)_{\ell}}{k!\,\ell!(c)_{k+\ell}} \, z^kw^{\ell} \ ,
\end{equation}
which is again convergent within $|z|<1$ and $|w|<1$ and then extended everywhere else by analytic continuation. The Appell function may be expanded in terms of ordinary hypergeometric functions through
\begin{eqnarray}
& &F_{1}(a,b_1,b_2,c|z,w) \,=\, \\
& &\sum\limits_{k=0}^{+\infty}\frac{(a)_{k}(b_1)_k(b_2)_{k}(c-a)_k}{k!\,(c+k-1)_{k}(c)_{2k}}z^kw^k{}_2F_{1}(a+k,b_1+k,c+2k|z){}_2F_{1}(a+k,b_2+k,c+2k|w) .\nonumber
\end{eqnarray}

 \bibliographystyle{utphys}
  \bibliography{references}
\end{document}